\documentclass{article}

\usepackage{PRIMEarxiv}

\usepackage[utf8]{inputenc} 
\usepackage[T1]{fontenc}    
\usepackage{hyperref}       
\usepackage{url}            
\usepackage{booktabs}       
\usepackage{amsfonts}       
\usepackage{nicefrac}       
\usepackage{microtype}      
\usepackage{lipsum}
\usepackage{fancyhdr}       
\usepackage{graphicx}       
\graphicspath{{media/}}     

\usepackage{float}
\restylefloat{table}
\usepackage{adjustbox}
\usepackage{multirow}
\usepackage[table,xcdraw]{xcolor}
\usepackage{subfig}
\usepackage{amsmath}

\title{Investigate The ESG Scoring Methodology
}

\author{
  Zhi Chen\\
  School of Business, Stevens Institute of Technology \\
  Hoboken, NJ, United States \\
  \texttt{\{zchen100\}@stevens.edu}
}

\begin{document}
\maketitle

\begin{abstract}
Whether the Refinitiv provide a reliable and trusted methodology in the process of aggregating 10 category scores to overall score?
Specifically, solving the first question will help us observe patterns of category weights of companies in various sectors, and reveal the different amount of attention assigned on each sector.
Looking at the weight changes year over year will help us understand how the Refinitiv reflect the historical and current environment. 
Finding the relationship leads to a possibility to create a predictive model of future ESG scores given a company's raw data. 
By comparing with the raw data and category's predictive ability, we can know whether the Refinitiv provide an useful and effective methodology to summary the data points. 
\end{abstract}


\section{Introduction}
Non-financial performance, especially Corporate Social Performance (CSP), has increasingly gained attention and importance among investors and firms. 
The corporate social performance of companies is often characterized through environmental, social, and governance (ESG) factors. 
Investors are incorporating ESG factors, alongside traditional financial metrics into the investment decision-making process. This integration is driven by the belief that ESG information is crucial in identifying companies that are potentially at risk of costly events linked to ESG issues, such as environmental damages or lawsuits stemming from corporate misconduct or fraud. These types of risks, which are not typically captured in traditional financial reports, can significantly impact a company's long-term financial stability and reputation. Hence, by considering ESG factors, investors aim to gain a more comprehensive understanding of a company's overall risk profile and sustainability.
Therefore, ESG investing aligns with the growing demand among investors to achieve a dual objective: to realize financial returns while also effectively managing risks that may arise from environmental, social, or governance shortcomings.

A company's ESG factors is often evaluated by the ESG ratings, which are developed by independent rating agencies. 
These agencies collect and assess a numbers of variables relating to a company's ESG practices. Subsequently, these variables are integrated into their distinct methodologies. The outcome is typically presented as a single score or rating. This score or rating acts as a concise summary of the company's ESG performance, offering investors a rapid and comparative perspective. 

However, the way the ESG ratings are calculated, by aggregating numbers of raw data into single number, may be detrimental to the prediction power of the ESG related variables collected. One possible explanation could be the introduction of noise during the aggregation process. This noise could stem from issues related to data quality and availability, where overlapping or incomplete data sets impact the accuracy of the final score. Another possible explanation could be the loss of important details. In the process of aggregation, specific and significant information about various ESG aspects could be overlooked or diluted, leading to a less precise reflection of a company's true ESG performance.

Additionally, there exists a notable divergence in the ratings issued for the same firm by different ESG rating providers. \cite{gibson2021esg} observed that the average pairwise correlation for the overall ESG ratings from the seven data provides is 45\%. This divergence contrasts sharply with the higher average correlations observed between credit ratings from Moody’s and S\&P. According to \cite{berg2022aggregate}, the correlation between these two credit ratings has reached 99\%. Discrepancies between ESG scores and the low correlation can be accounted by the lack of clear consensus around what should be measured when putting together an ESG rating \cite{clement2022improving}. 

Therefore, recognizing the potential limitations and issues associated with the use of aggregated ESG ratings, the object of this paper is to examine the underlying raw variables, which are utilized in calculating the aggregated scores. We aim to assess and compare the predictive capabilities of these raw variables versus the aggregated ESG scores. This comparison will focus on determining which approach (raw variables or aggregated scores) offers a more accurate and reliable prediction of company stock returns and volatility. Further, we aim to identify which specific raw variables are most significant in the prediction.

\subsection{ESG-related Risk} \label{section: ESG-related Risk}

ESG-related risks can manifest in various ways.
Climate change has begun to significantly impact a large number of firms, leading to direct financial losses \cite{gold2019pg}. 
Utility sector companies will be increasingly exposed to threats stemming from hurricanes, rising sea levels, and other climate-related events. For example, in January 2019, Pacific Gas \& Electric (PG\&E) declared bankruptcy. It was due to the massive financial liabilities from wildfires in California that were linked to the company's maintenance of power lines. 
Energy sector companies, according to Intergovernmental Panel on Climate Change (IPCC), is also vulnerable to climate change \cite{edenhofer2015climate}. For example, temperature variability is likely to influence energy demand via impacts on heating and cooling requirements, thereby rendering energy enterprises susceptible to price volatility and supply disruptions \cite{yalew2020impacts}.

ESG regulations, while aimed at promoting sustainability and responsible practices, may concurrently pose various risks to companies \cite{plastun2020sdgs}. Failure to comply with ESG regulations can result in fines, penalties and legal liabilities. On the other hand, implementing the necessary changes to comply with new or existing ESG regulations can be costly. This may include investments in new technologies, infrastructure, or personnel training to meet environmental standards or improve governance structures. ESG regulations may also extend to a company's supply chain, requiring not only the company but also its suppliers to adhere to certain standards. This can create challenges in managing supply chain relationships.

A company's reputation is a crucial intangible asset that can significantly impact its performance \cite{gatzert2015impact}. Investors and shareholders may lose confidence in a company with a bad reputation, which can affect the share price negatively and hinder the ability to raise capital. Take a look at a scandal from German auto manufacturer Volkswagen \cite{blackwelder2016volkswagen}.
In 2015-09 the U.S.'s Environmental Protection Agency (EPA) accused the Volkswagen violating the Clean Air Act.
The company engineered its vehicles to circumvent government emissions testing protocols. This resulted in an immediate drop in share price followed by lawsuits.
The Volkswagen scandal is evidence of the need for greater corporate transparency.
The scandal also highlights the failure of traditional valuation models, such as discounted cash flow, to capture the full range of risk companies face today.
It shows the potential benefits of assessing companies with non-financial data that reflect ESG risk.

There are a series of ESG-related risk that can influence company's performance. This give us a hint that ESG encompass a wealth of information that can significantly aid in mitigating risks, which solely examining traditional financial statements may not achieve. This is because ESG criteria delve into a company's broader contextual operations, shedding light on its environmental impact, social relationships, and governance structures. Analyzing ESG factors provides a holistic understanding of a company’s long-term sustainability which is vital for investment decision-making.

\subsection{ESG Rating} \label{section: ESG Rating}
In order to provide the transparency of a company's policies and efforts concerning sustainability, rating agencies issue ESG rating. There are multiple data providers such as (1) ASSET4 by Thomson Reuters, (2) Kinder Lydenberg Domini \& Co. (KLD) by
MSCI and Sustainability Asset Management
Group (SAM), (3) S\&P Global Ratings, and Bloomberg.

Typically, an ESG rating is presented as a single number or letter, serving as a concise indicator of a company's overall ESG management and performance. This singular rating is aggregated hierarchically using numerous raw variables, following a pyramid-like structure. At the base of this pyramid are the raw data points. These data points, encompassing a wide range of factors from carbon emissions and labor practices to board diversity and corporate ethics, are 
synthesized at each ascending level of the pyramid-like rating system.

In the aggregation process of determining an ESG rating, some ESG raters apply a percentile ranking methodology. This approach involves comparing a company's ESG performance against that of its peers. For example, if a company receives an ESG score of 80, it indicates that its ESG performance surpasses that of 80\% of its peers. According to ASSET4 Professional Guide \cite{reuters2011asset4}, all scores are normalized by using z-scoring and benchmarked against the complete universe of 4000 companies. As per the Bloomberg ESG score documentation, the overall ESG score of a company is determined based on its percentile ranking. E, S score are industry-specific and governance scores is country-specific. 

The percentile ranking system in ESG evaluations offers valuable insights into a company's position relative to its peers. However, they do not directly reveal the absolute performance or the specific ESG metrics achieved. For a more comprehensive understanding of a company's ESG profile, examining the actual values, such as specific emissions levels, diversity ratios, or governance scores, is crucial. 

Having discussed the structure and methodology of the ESG ratings, we now begin to consider the sources and nature of the underlying data points.
These data points can be sourced from a variety of documents. Some common sources of raw data for ESG ratings are: company disclosures, government and regulatory filings, survey and questionnaires, non-governmental organization (NGO) reports, media and news. 
Among of these sources, corporate disclosures, principally in the form of Corporate Social Responsibility (CSR) reports, serve as the primary resource. CSR reports provide detailed information on a company's ESG practices and performance, such as environmental stewardship, labor and community relations, governance structures, and ethical conduct. Many companies publish CSR reports on an annual basis. The issuance of CSR reports can be at any time during the year. The EU Non-Financial Reporting Directive (NFRD) mandates certain large companies to disclose non-financial and diversity information \cite{cuomo2022effects}.

Several organizations guide the publication of CSR reports for companies. For example, Global Reporting Initiative (GRI) provides a comprehensive framework to enable company to understand and report on their impacts on the economy, environment and people in a comparable and credible way, thereby increasing transparency on their contribution to sustainable development \cite{vartiak2016csr}.

Despite these rating agencies take the similar resource, the ratings assigned to companies by different rating agencies differ noticeably.
\cite{semenova2015validity} investigate the convergence of the environmental rating provided by KLD, ASSET4 and Global Engagement Services (GES) for US companies from the MSCI World universe (2003-2011). The study shows that the correlations between ratings are 25\% on average. 
\cite{chatterji2014ratings} examine correlations between overall ESG data of ASSET4 and KLD as well as other rating agencies. The results shows the correlations range from 13\% to 52\%, which indicates low convergence among raters. 

\cite{dorfleitner2015measuring} compares different rating approaches using ESG overall scores and individuals E, S, G scores. The results suggests an evident lack in the convergence of ESG measurement. The E,S,G scores of the ASSET4 and Bloomberg exhibit positive correlations between 0.47 and 0.60. The correlation of overall ESG score between these two raters is 0.62. 
In summary, the ratings approaches lead to distinct results. 

The divergence primarily stems from not only the differing emphases on the underlying data but also variations in how these raw variables are aggregated \cite{berg2022aggregate}. This lack of uniformity among data providers suggests that relying solely on these aggregated ratings might not be the most appropriate approach for assessing a company's ESG performance. Instead, a more reliable method would be to directly analyze the raw data. This approach not only circumvents the biases and methodological discrepancies inherent in aggregated ratings but also provides a more detailed understanding of a company's ESG profile.

\subsection{Question to be Addressed}
\label{section: Question to be Addressed}

This section investigates the top of Refinitiv ESG scoring methodology pyramid: starting from 10 ESG category scores to ESG overall score. The analysis will help address the following questions: 

\begin{itemize}
    \item How 10 category scores are weighted for different sectors?
    \item How do the weights change over time?
    \item How 10 category scores are weighted for companies within the same sector?
    \item What is the relationship between category score and overall ESG score?
    \item What is the predictive ability of category score to stock performance?
\end{itemize}






\section{Data}

\subsection{ASSET4 Data and Its Structure}
\label{section: Data}

The ASSET4 data initiated in 2002 and now provided ESG scores of over 4300 companies worldwide. The coverage universe
comprises listed companies ranging from the
S\&P 500, Russell 1000, MSCI Europe, FTSE
250, ASX 300 to the MSCI World Index and
the 250 MSCI Emerging Markets companies \cite{dorfleitner2015measuring}.


The ASSET4 data has received endorsements from several reputable organizations, reflecting its reliability and significance in the field.
ASSET4 is signatory of the UN's Principles for Responsible Investment and a contributing member to organizations and initiatives like: UNEP-FI, Eurosif, UKSIF, USSIF, and the Veres Coalition \cite{ribando2010new}.  

\begin{figure}[H]
\caption{Data Structure of ASSET4}
\label{Data Structure of ASSET4}
\centering
\includegraphics[width=1\textwidth]{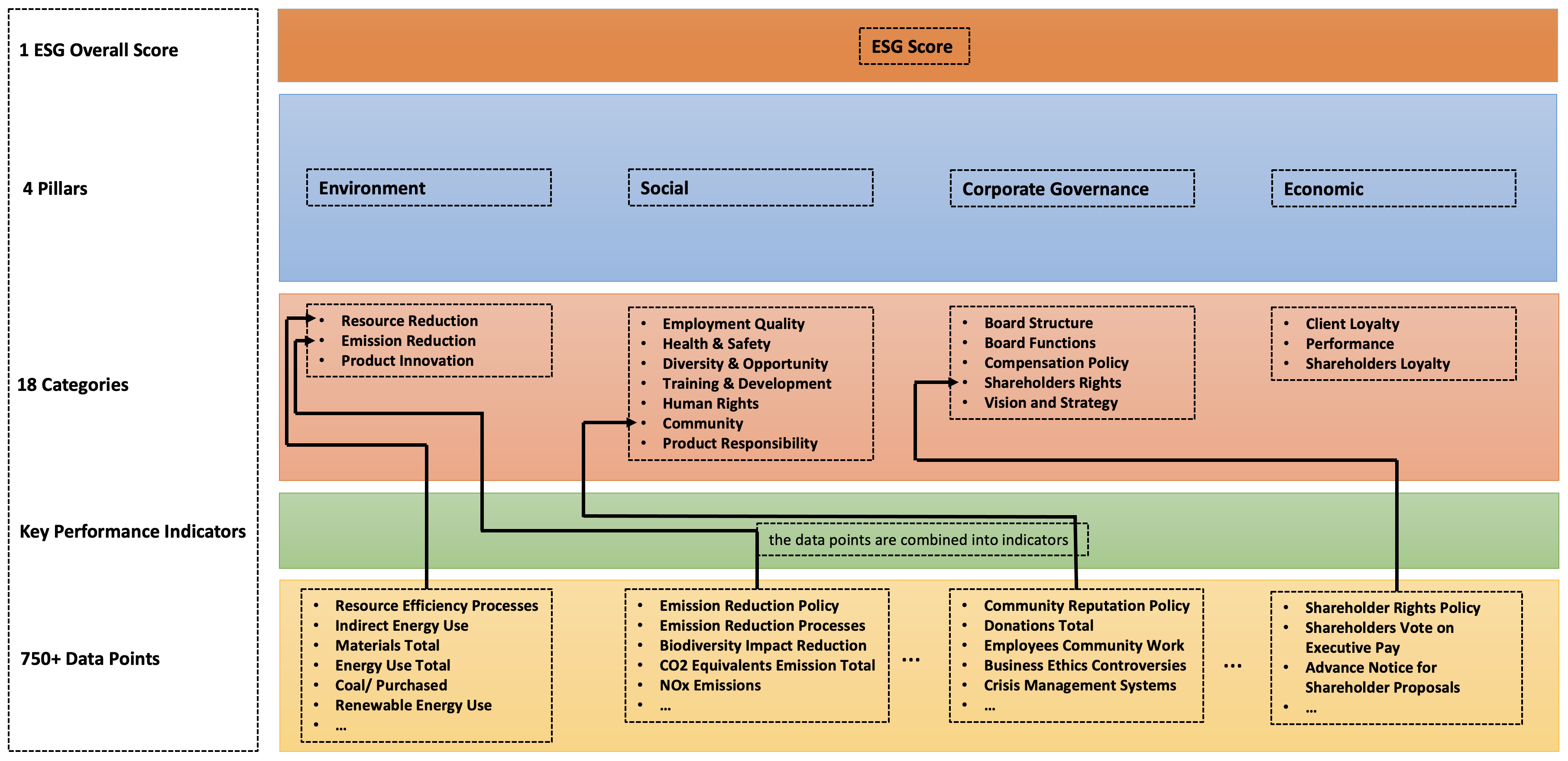}
\end{figure}

ASSET4 uses a hierarchical, four-level rating system for evaluating each company's ESG performance. See the figure \ref{Data Structure of ASSET4}. On the left-hand side, it presents an overview of the structure, labeling each layer with its respective name. On the right-hand side of the diagram it displayed the specific items contained within each of these layers.

\begin{itemize}
    \item Overall ESG Score (Fifth Level) \\
    This is the apex of the pyramid, where a composite ESG score is calculated for each company, providing a single number (between 0 and 100). A high (low) score indicates a strong (poor) company performance.
    
    The overall score is based on scores in four key pillars: Environmental, Social, Corporate Governance, and Economic Performance.
    \item 4 Pillars (Fourth Level) \\
    Each pillar score is calculated from its own several category scores. For example, environmental pillar has three category: resource reduction, emission reduction, product innovation. Therefore environmental pillar scores is based on these three category scores. 
    
    \item 18 Category Scores (Third Level) \\
    Each category is comprehensively defined by its corresponding indicators and data points. For instance, the 'Resource Reduction' category is computed through approximately 60 data points. These data points collectively offer a multifaceted view and capture various aspects of a company's efforts and performance in resource reduction. Importantly, these data points are exclusive to their respective categories, ensuring that each data point is unique to a single category and there is no overlap. The relationship is depicted in the figure \ref{Data Structure of ASSET4} through a black arrow line.
    
    \item Key Performance Indicator (KPI) (Second Level) \\
    These indicators are computed from combination of data points. The ASSET4 define the rules of how to determine the KPI using a group of data points. For example, to determine an indicator 'resource reduction/Monitoring', it need to check 4 data points. The rule for this indicator might be formulated as follows: if data point 1 and data point 5 both equal 'yes', or if data point 3 equals 'yes', or data point 4 equals 'no', then the indicator is assigned a 'yes'. Otherwise, it is assigned a 'no'. 
    
    \item Raw Data Points (First Level) \\
    At the base of the ASSET4 data structure, there are around 700 data points.
    One of the reasons we use this dataset is they supply raw ESG data that could be used by us to devise our own measurement.

\end{itemize}

It is important to note that the ASSET4's methodology and data structure partially changed in 2017 \cite{ThomsonReuters}. Before 2017, ASSET4 comprised four pillars: (1) environmental pillar, (2) social pillar, (3) corporate governance pillar and (4) economic pillar. In 2017, the economic pillar was removed, leaving three pillars. To maintain consistency in data structure and ensure comparability of data over time, we have chosen to focus solely on the three remaining pillars in our analysis.

Our sample consists of stocks that are constituents of S\&P500 and Russell 2000. The S\&P500 includes 500 of the largest U.S. companies. The Russell 2000, on the other hand, focuses on 2000 smaller business. Together, these indexes provide a comprehensive view of the U.S. stock market, representing both large-cap and small-cap segments. 

Among this stock universe, we specifically focus on the energy-related companies. The reasons are as follows:
\begin{itemize}
    \item ESG sensitivity \\
    The energy sector has a strong environmental impact, primarily due to greenhouse gas emissions, resource extraction processes, and waste management. Further, as the world moves towards sustainable energy solutions to combat climate change, energy companies are at the forefront of this transition. Thus, ESG scores are more relevant to energy sector. 
    
    \item Regulatory and policy scrutiny \\
    Energy companies often face stringent regulations due to their environmental impact. Focusing on them can help understand how these companies are managing regulatory risks and complying with evolving environmental laws.
    
\end{itemize}

We determined the energy-related companies based on an industry classfification system operated by Refinitiv: The Refinitiv Business Classification (TRBC). According to TRBC, there are 51 energy-related companies in the combined stock universe comprised by S\&P500 and Russell 2000. 

For each company in energy sector, we collect different levels of data: ESG overall score, pillar score, category score, raw data point with data ranging from 2002-2020.  

\section{Methodology}

We refined the data to the companies that are included in the Standard \& Poor's 500 and Rusell 2000 with data ranging from 2002-2020. Among of these companies, we use The Refinitiv Business Classification (TRBC) (i.e., an industry classification system that is owned and operated by Refinitiv.) to break down correlated companies and organizations. 

We focus on energy, Banking \& Investment Services, Technology Equipment sector, whose TRBC hierarchical ID is 5010, 5510, 5710 respectively. After cross checking the list, there are in total 48, 157, 74  S\&P 500 and Rusell 2000 companies that belongs to these three sectors respectively. 

\begin{itemize}
    \item Subset the regression data by sector \\
    Select observations of all companies within the same sector in a given year. Regress overall score on 10 category scores.
    \item Subset the regressions data by year range \\
    Select observations of all companies within the same sector from 2002 to 2020. Regress overall score on 10 category scores.
    \item Subset the regressions data by individual company \\
    Select observations of one company of a given sector from 2002 to 2020. Regress overall score on 10 category scores.
    \item Regress stock performances on 10 category scores 
    \item Regress stock performances on each of 10 category scores
\end{itemize}

\section{Result}

\subsection{Subset the regression data by sector}

Table \ref{tab: R-squared for different sectors} displays the adjusted R-squared of each model when run regressions for different sectors in a certain year. All the results are very close to 1 showing the models are perfect fitting. It indicates that the Refinitiv will assign the same weights on companies within the same business classification every year.

Table \ref{tab: Category weights of different sectors in 2020} displays the category weights of different sectors in 2020. The columns list various sectors while each row signifies the weights of those sector for a certain category. Figure \ref{fig: Category Weight 2020} visualizes the distribution of category weight that Refinitiv assign on various sectors.

\begin{table}[htp]
  \centering
  \caption{R-squared for different sectors}
    \begin{adjustbox}{width=1.2\columnwidth,center}
    \begin{tabular}{|c|r|r|r|r|r|r|r|r|}
    \toprule
          & \multicolumn{1}{c|}{\textbf{Enery}} & \multicolumn{1}{c|}{\textbf{Healthcare}} & \multicolumn{1}{c|}{\textbf{Banking}} & \multicolumn{1}{c|}{\textbf{Industrial Goods}} & \multicolumn{1}{c|}{\textbf{Food \& Beverages}} & \multicolumn{1}{c|}{\textbf{Technology Equipments}} & \multicolumn{1}{c|}{\textbf{software \& IT Services}} & \multicolumn{1}{c|}{\textbf{Utilities}} \\
    \midrule
    \textbf{2002} & 0.99697 & 0.998609 & 0.983256 & 0.974103 & 0.999296 & 0.997593 & 1     & 0.990682 \\
    \midrule
    \textbf{2003} & 0.996206 & 0.998715 & 0.954447 & 0.991258 & 0.99989 & 0.999175 & 1     & 0.993896 \\
    \midrule
    \textbf{2004} & 0.98098 & 0.997919 & 0.964815 & 0.984316 & 0.999706 & 0.964676 & 1     & 0.996543 \\
    \midrule
    \textbf{2005} & 0.993732 & 0.998685 & 0.994572 & 0.991786 & 0.999569 & 0.979719 & 1     & 0.997413 \\
    \midrule
    \textbf{2006} & 0.996927 & 0.998217 & 0.988085 & 0.989336 & 0.999378 & 0.981429 & 1     & 0.998229 \\
    \midrule
    \textbf{2007} & 0.992514 & 0.996386 & 0.973451 & 0.991273 & 0.998586 & 0.976454 & 1     & 0.997392 \\
    \midrule
    \textbf{2008} & 0.995306 & 0.996528 & 0.986566 & 0.989284 & 0.999376 & 0.992941 & 1     & 0.996729 \\
    \midrule
    \textbf{2009} & 0.993802 & 0.993319 & 0.978081 & 0.995376 & 0.999432 & 0.992372 & 1     & 0.99833 \\
    \midrule
    \textbf{2010} & 0.996713 & 0.991681 & 0.985389 & 0.994538 & 0.999814 & 0.985332 & 1     & 0.997959 \\
    \midrule
    \textbf{2011} & 0.992707 & 0.991859 & 0.989177 & 0.993893 & 0.998897 & 0.97931 & 1     & 0.997637 \\
    \midrule
    \textbf{2012} & 0.993843 & 0.993246 & 0.987988 & 0.989548 & 0.999164 & 0.991275 & 1     & 0.997054 \\
    \midrule
    \textbf{2013} & 0.994022 & 0.994591 & 0.982628 & 0.984087 & 0.999275 & 0.9952 & 1     & 0.995258 \\
    \midrule
    \textbf{2014} & 0.994814 & 0.992892 & 0.979481 & 0.992303 & 0.997289 & 0.993955 & 1     & 0.995488 \\
    \midrule
    \textbf{2015} & 0.996166 & 0.995816 & 0.995553 & 0.993772 & 0.999679 & 0.98432 & 1     & 0.996976 \\
    \midrule
    \textbf{2016} & 0.996499 & 0.997318 & 0.996128 & 0.994721 & 0.999922 & 0.987076 & 1     & 0.997724 \\
    \midrule
    \textbf{2017} & 0.997623 & 0.997246 & 0.996463 & 0.995371 & 0.999931 & 0.979785 & 1     & 0.997959 \\
    \midrule
    \textbf{2018} & 0.994751 & 0.997903 & 0.996837 & 0.99693 & 0.999938 & 0.985486 & 1     & 0.998783 \\
    \midrule
    \textbf{2019} & 0.992809 & 0.997995 & 0.998374 & 0.996757 & 0.999935 & 0.985255 & 1     & 0.99854 \\
    \midrule
    \textbf{2020} & 0.995628 & 0.997911 & 0.998581 & 0.996816 & 0.999886 & 0.984331 & 1     & 0.999316 \\
    \bottomrule
    \end{tabular}%
  \label{tab: R-squared for different sectors}%
  \end{adjustbox}
\end{table}%

\begin{table}[htbp]
  \centering
  \caption{Category weights of different sectors in 2020}
  \begin{adjustbox}{width=1.2\columnwidth,center}
    \begin{tabular}{|l|r|r|r|r|r|r|r|r|r|r|}
    \toprule
          & \multicolumn{1}{l|}{Resource Use Score} & \multicolumn{1}{l|}{Emissions Score} & \multicolumn{1}{l|}{Environmental Innovation Score} & \multicolumn{1}{l|}{Workforce Score} & \multicolumn{1}{l|}{Human Rights Score} & \multicolumn{1}{l|}{Community Score} & \multicolumn{1}{l|}{Product Responsibility Score} & \multicolumn{1}{l|}{Management Score} & \multicolumn{1}{l|}{Shareholders Score} & \multicolumn{1}{l|}{CSR Strategy Score} \\
    \midrule
    Energy - Fossil Fuels & 0.1248 & 0.1727 & 0.0712 & 0.1163 & 0.1645 & 0.1001 & 0.0601 & 0.1688 & 0.0479 & -0.0100 \\
    \midrule
    Healthcare Services \& Equipment & 0.0707 & 0.0585 & 0.0332 & 0.0730 & 0.1158 & 0.1335 & 0.1396 & 0.2383 & 0.0723 & 0.0550 \\
    \midrule
    Banking \& Investment Services & 0.0622 & 0.0176 & 0.0968 & 0.1937 & 0.0664 & 0.1185 & 0.0863 & 0.2424 & 0.0750 & 0.0334 \\
    \midrule
    Industrial Goods & 0.0812 & 0.0825 & 0.1666 & 0.0941 & 0.1066 & 0.0836 & 0.0865 & 0.1978 & 0.0629 & 0.0367 \\
    \midrule
    Food \& Beverages & 0.1310 & 0.1276 & 0.0270 & 0.1130 & 0.1247 & 0.0839 & 0.1305 & 0.1755 & 0.0536 & 0.0356 \\
    \midrule
    Technology Equipment & 0.0730 & 0.1033 & 0.1099 & 0.1222 & 0.1170 & 0.1002 & 0.0922 & 0.1919 & 0.0718 & 0.0201 \\
    \midrule
    Software \& IT Services & 0.0463 & 0.0308 & 0.0617 & 0.0694 & 0.0617 & 0.1542 & 0.1131 & 0.3085 & 0.0925 & 0.0617 \\
    \midrule
    Utilities & 0.1313 & 0.1530 & 0.1223 & 0.1315 & 0.0718 & 0.1052 & 0.0589 & 0.1710 & 0.0386 & 0.0314 \\
    \bottomrule
    \end{tabular}%
  \label{tab: Category weights of different sectors in 2020}%
  \end{adjustbox}
\end{table}%

\subsection{Subset the regressions data by year range}

Table \ref{tab: Adj.R-squared for different sectors over time} displays the adjusted r-squared for a sector over time.
We found that adjusted R-squared of all regression models are very close or equal to 1. It indicates that the Refinitiv assigned the same weights from year to year. 

By checking the Refinitiv document, it turns out when update happens such as including more companies or changing methodology, Refinitiv will recalculate all the previous year's scores with the new weights. This would be good for comparison to see how the companies have grown under the same standards.

\begin{table}[htbp]
  \centering
  \caption{Adj.R-squared for different sectors over time}
  \begin{adjustbox}{width=1.2\columnwidth,center}
    \begin{tabular}{|l|r|r|r|r|r|r|r|r|}
    \toprule
          & \multicolumn{1}{c|}{\textbf{Enery}} & \multicolumn{1}{c|}{\textbf{Healthcare}} & \multicolumn{1}{c|}{\textbf{Banking}} & \multicolumn{1}{c|}{\textbf{Industrial Goods}} & \multicolumn{1}{c|}{\textbf{Food \& Beverages}} & \multicolumn{1}{c|}{\textbf{Technology Equipments}} & \multicolumn{1}{c|}{\textbf{software \& IT Services}} & \multicolumn{1}{c|}{\textbf{Utilities}} \\
    \midrule
    Adjusted R-squared & 0.99  & 1.00  & 0.99  & 0.99  & 1.00  & 0.99  & 1.00  & 1.00 \\
    \bottomrule
    \end{tabular}%
  \label{tab: Adj.R-squared for different sectors over time}%
  \end{adjustbox}
\end{table}%

\subsection{Subset the regressions data by individual company}

Table \ref{tab: Category weights of individual company in energy sector over time}, table \ref{tab: Category weights of individual company in financial sector over time} display the regression coefficients of part of companies in energy sector, financial sector, respectively. In the last row these tables list the r-squared of corresponding regression. 
We found that typically there is one pattern of weights for companies within the same sector, which is making sense as we conclude in the above experiments that the Refinitiv assign the same weights for companies within the same sector. But we also found some outliers whose weights are not satisfied the pattern. This is because the number of observations of those companies are not enough to run a valid regression, then we just get an abnormal coefficients.

\begin{table}[htbp]
  \centering
  \caption{Category weights of individual company within energy sector over time}
  \begin{adjustbox}{width=1.2\columnwidth,center}
    \begin{tabular}{|c|r|r|r|r|r|r|r|r|}
    \toprule
          & \multicolumn{1}{c|}{\textbf{SLB}} & \multicolumn{1}{c|}{\textbf{HAL}} & \multicolumn{1}{c|}{\textbf{DEN}} & \multicolumn{1}{c|}{\textbf{NBR}} & \multicolumn{1}{c|}{\textbf{OKE}} & \multicolumn{1}{c|}{\textbf{13123810}} & \multicolumn{1}{c|}{\textbf{CPE}} & \multicolumn{1}{c|}{\textbf{STNG}} \\
    \midrule
    \textbf{CSR Strategy Score} & 0.037 & 0.037 & 0.037 & 0.037 & 0.037 & 0.080242565 & 0.08  & 0.089 \\
    \midrule
    \textbf{Community Score} & 0.093 & 0.093 & 0.093 & 0.093 & 0.093 & 0.061501444 & 0.062 & 0.077 \\
    \midrule
    \textbf{Emissions Score} & 0.149 & 0.149 & 0.149 & 0.149 & 0.149 & 0.106787988 & 0.107 & 0.075 \\
    \midrule
    \textbf{Environmental Innovation Score} & 0.046 & 0.046 & 0.046 & 0.046 & -0    & 3.21652E-18 & 3E-18 & -0 \\
    \midrule
    \textbf{Human Rights Score} & 0.149 & 0.149 & 2E-17 & 0.149 & 0.149 & 0     & 0     & 0.21 \\
    \midrule
    \textbf{Management Score} & 0.121 & 0.186 & 0.186 & 0.186 & 0.186 & 0.151456572 & 0.151 & 0.149 \\
    \midrule
    \textbf{Product Responsibility Score} & 0.043 & 0.043 & 0.043 & 0.043 & 0.043 & -0.009397837 & -0.01 & 0.063 \\
    \midrule
    \textbf{Resource Use Score} & 0.13  & 0.13  & 0.13  & 0.13  & 0.13  & 0.063833602 & 0.064 & 0.045 \\
    \midrule
    \textbf{Shareholders Score} & 0.121 & 0.056 & 0.056 & 0.056 & 0.056 & 0.05133956 & 0.051 & 0.042 \\
    \midrule
    \textbf{Workforce Score} & 0.111 & 0.111 & 0.111 & 0.111 & 0.111 & 0.128334462 & 0.128 & 0.041 \\
    \midrule
    \textbf{const} & -0    & -0    & 2E-16 & -0    & 7E-15 & 0.030512217 & 0.031 & 0.029 \\
    \midrule
    \textbf{r-squared} & 1     & 1     & 1     & 1     & 1     &   NA    &   NA    &  NA \\
    \bottomrule
    \end{tabular}%
  \label{tab: Category weights of individual company in energy sector over time}%
  \end{adjustbox}
\end{table}%

\begin{table}[htbp]
  \centering
  \caption{Category weights of individual company in financial sector over time}
  \begin{adjustbox}{width=1.2\columnwidth,center}
    \begin{tabular}{|c|r|r|r|r|r|r|}
    \toprule
          & \multicolumn{1}{c|}{\textbf{AXP}} & \multicolumn{1}{c|}{\textbf{CMA}} & \multicolumn{1}{c|}{\textbf{BAC}} & \multicolumn{1}{c|}{\textbf{HBAN}} & \multicolumn{1}{c|}{\textbf{ASO}} & \multicolumn{1}{c|}{\textbf{AROW}} \\
    \midrule
    \textbf{CSR Strategy Score} & 0.048 & 0.048 & 0.048 & 0.048 & 0     & 0 \\
    \midrule
    \textbf{Community Score} & 0.12  & 0.12  & 0.12  & 0.12  & 0.161082 & 0.1405 \\
    \midrule
    \textbf{Emissions Score} & 0.024 & 0.024 & 0.024 & 0.024 & -8.2E-19 & 5.92E-17 \\
    \midrule
    \textbf{Environmental Innovation Score} & 0.096 & 0.096 & 0.096 & 0.096 & 0     & 1.45E-17 \\
    \midrule
    \textbf{Human Rights Score} & 0.096 & 0.096 & 0.096 & 0.096 & 0     & 0 \\
    \midrule
    \textbf{Management Score} & 0.24  & 0.24  & 0.24  & 0.24  & 0.198504 & 0.241024 \\
    \midrule
    \textbf{Product Responsibility Score} & 0.088 & 0.088 & 0.088 & 0.088 & 0     & 0.037353 \\
    \midrule
    \textbf{Resource Use Score} & 0.024 & 0.024 & 0.024 & 0.024 & 3.76E-18 & 2.86E-16 \\
    \midrule
    \textbf{Shareholders Score} & 0.072 & 0.072 & 0.072 & 0.072 & 0.061479 & 0.061338 \\
    \midrule
    \textbf{Workforce Score} & 0.192 & 0.192 & 0.192 & 0.192 & 0.202207 & 0.188658 \\
    \midrule
    \textbf{const} & 4.44E-16 & -6.7E-16 & -3.8E-15 & 9.16E-16 & 0.014223 & 0.007013 \\
    \midrule
    \textbf{r-squared} & 1     & 1     & 1     & 1     &   NA    & NA  \\
    \bottomrule
    \end{tabular}%
  \label{tab: Category weights of individual company in financial sector over time}%
  \end{adjustbox}
\end{table}%

\section{Conclusion}
For the individual regressions of the raw variables it appears that Refinitiv weighs certain raw variables as more important depending on a company’s sector/industry. For example, a technology company’s CO2 emissions will weigh less on their Environmental score compared to an energy production company’s score. There are also various raw variables that are still collected but remain inactive which means they are not included in the final calculations that eventually accumulate to the total ESG score.
Overall looking at the raw variable data it seems that Refinitiv have not changed the calculation methods of many of the companies since 2002. This is reflected by an r-squared value of 1 or very close to 1 for many of the regressions. This would be very strange since the world and domestic economies have vastly changed as well as the environmental, social, and governance principles in which companies are judged. For this reason, there is also another explanation for the same calculations over the past 20 years. It can also be that when Refinitiv releases new scores with new weights. They then recalculate all the previous year’s scores with the new weights. This would be good for comparison to see how the companies have grown under the same standards however, when not specifically stated can lead to misinformation and misuse of the data.


\bibliographystyle{unsrt}  
\bibliography{references}

\end{document}